\documentclass{article}

\pdfoutput=1

\usepackage{titlesec}
\usepackage{caption}
\usepackage{subcaption}
\usepackage{booktabs}
\usepackage{multirow}
\usepackage{multicol}
\usepackage{xspace}
\usepackage{dsfont}
\usepackage{bbm}
\usepackage{bm}
\usepackage{pifont}
\usepackage{amsmath}
\usepackage{amsthm}
\usepackage{mathtools}
\usepackage{stackengine,scalerel}
\usepackage{amssymb}
\usepackage{enumitem}
\usepackage{wrapfig}
\usepackage{caption}
\usepackage{cancel}
\setlist{leftmargin=5mm}

\usepackage{hyperref}
\usepackage[capitalise, noabbrev]{cleveref}

\titleformat{\paragraph}[runin]{\bfseries\itshape}{\theparagraph}{1em}{}
\titlespacing*{\paragraph}{0pt}{3.25ex plus 1ex minus .2ex}{1em}

\makeatletter \renewcommand\paragraph{\@startsection{paragraph}{4}{\z@} {2mm \@plus1ex \@minus.2ex} {-0.7em} {\normalfont\normalsize\itshape}} \makeatother

\newcommand{\listheader}[1]{\item \emph{#1} }

\newcommand\overstar[1]{\ThisStyle{\ensurestackMath{%
  \setbox0=\hbox{$\SavedStyle#1$}%
  \stackengine{0pt}{\copy0}{\kern.2\ht0\smash{\SavedStyle*}}{O}{c}{F}{T}{S}}}}

\newtheorem{theorem}{Theorem}[section]

\newtheorem{proposition}[theorem]{Proposition}

\newenvironment{customTheorem}[1]{%
  \renewcommand\thetheorem{#1}
  \proposition
}{\endtheorem}

\newcommand{\ie}[0]{\emph{i.e.}\xspace}
\newcommand{\etc}[0]{\emph{etc.}\xspace}
\newcommand{\eg}[0]{\emph{e.g.}\xspace}
\newcommand{\vs}[0]{\emph{vs.}\xspace}

\newcommand{\dataset}[0]{$\mathcal{D}$\xspace}

\newcommand{\sampler}{\textsc{SVP-CF}\xspace}
\newcommand{\samplerprop}{\textsc{SVP-CF-Prop}\xspace}

\newcommand{\EE}{\operatornamewithlimits{\mathbb{E}}} 

\usepackage{hyperref}

\usepackage[accepted]{icml2021}

\icmltitlerunning{SVP-CF: Selection via Proxy for Collaborative Filtering Data}

\begin{document}

\twocolumn[
\icmltitle{SVP-CF: Selection via Proxy for Collaborative Filtering Data}



\icmlsetsymbol{equal}{*}

\begin{icmlauthorlist}
\icmlauthor{Noveen Sachdeva}{ucsd}
\icmlauthor{Carole-Jean Wu}{fb}
\icmlauthor{Julian McAuley}{ucsd}
\end{icmlauthorlist}

\icmlaffiliation{ucsd}{University of California, San Diego, CA, USA}
\icmlaffiliation{fb}{Facebook, Menlo Park, CA, USA}

\icmlcorrespondingauthor{Noveen Sachdeva}{nosachde@ucsd.edu}

\icmlkeywords{Machine Learning, ICML}

\vskip 0.3in
]



\printAffiliationsAndNotice{}  

\begin{abstract}
We study the practical consequences of dataset sampling strategies on the performance of recommendation algorithms. Recommender systems are generally trained and evaluated on \emph{samples} of larger datasets. Samples are often taken in a na\"ive or ad-hoc fashion: \eg by sampling a dataset randomly or by selecting users or items with many interactions. As we demonstrate, commonly-used data sampling schemes can have significant consequences on algorithm performance---masking performance deficiencies in algorithms or altering the \emph{relative} performance of algorithms, as compared to models trained on the complete dataset. Following this observation, this paper makes the following main contributions: (1) \emph{characterizing} the effect of sampling on algorithm performance, in terms of algorithm and dataset characteristics (\eg sparsity characteristics, sequential dynamics, \etc); and (2) designing \sampler, which is a data-specific sampling strategy, that aims to preserve the relative performance of models after sampling, and is especially suited to long-tail interaction data. Detailed experiments show that \sampler is more accurate than commonly used sampling schemes in retaining the relative ranking of different recommendation algorithms.
\end{abstract}

\section{Introduction} 
Representative \emph{sampling} of collaborative filtering (CF) data is a crucial problem from numerous stand-points and is generally performed at various levels: (1) mining hard-negatives while training complex recommendation algorithms over massive datasets \cite{eclare, sampling_cf_nn}; (2) down-sampling the item-space to estimate expensive ranking metrics \cite{sampled_metrics}; and (3) sub-sampling the entire dataset for reasons like easy-sharing, fast-experimentation, mitigating the significant environmental footprint of training resource-hungry machine learning models \cite{google_emissions, facebook_emissions, green_ai}, \etc In this paper, we are interested in finding a sub-sample of a dataset which has minimal effects on model utility evaluation \ie an algorithm performing well on the sub-sample should also perform well on the original dataset.

Preserving \emph{exactly} the same levels of performance on sub-sampled data over metrics like MSE, AUC, \etc is a very challenging problem. However, a simpler albeit useful problem is accurately preserving the \emph{ranking} or relative performance of different algorithms on sub-sampled data. For \eg, a sampling scheme that has a very low bias but high variance in preserving metric performance values has a lesser utility than a different sampling scheme with high amounts of bias but low variance, since the overall algorithm ranking is still preserved.

Performing careless and ad-hoc sampling such as randomly removing interactions, or making dense subsets by removing users/items with few interactions \cite{sigir20} can have adverse downstream repercussions. For \eg, sampling only the head-portion of a dataset, from a fairness and inclusion perspective, is inherently biased against minority-groups and benchmarking algorithms on this biased data is highly likely to propagate the original sampling biases. On the other hand, from an entirely performance view-point, accurately retaining the relative performance of different recommendation algorithms on much smaller sub-samples is a challenging research problem in itself.

Two prominent directions towards representative sampling of CF data are: (1) designing principled sampling strategies, especially for user-item interaction data; and (2) analyzing the performance of different sampling strategies, in order to better grasp which sampling scheme works ``better'' for which type of data. \emph{In this paper,} we explore both of these directions through the lens of expediting the recommendation algorithm development cycle, by:
\begin{itemize} \setlength\itemsep{0.1em}
    \item \emph{Characterizing} the efficacy of \emph{sixteen} different sampling schemes in accurately benchmarking various kinds of recommendation algorithms on smaller sub-samples.

    \item Proposing a data-specific sampling strategy, \sampler, which can dynamically sample the ``toughest'' portion of a CF dataset. \sampler is also specifically designed to handle the inherent data heterogeneity and missing-not-at-random properties in user-item interaction data.
    
\end{itemize}
%
Ultimately, our experiments reveal that \sampler outperforms all other sampling strategies and can accurately benchmark recommendation algorithms with roughly $50\%$ of the original data
leading to roughly $1.8\times$ time speedup.

\section{Related Work} \paragraph{Sampling CF data.} Sampling in CF-data has been a popular choice for three major scenarios. Most prominently, sampling is used for mining hard-negatives while training recommendation algorithms. Some popular approaches include randomly sampling negative interactions; using the graph-structure to find the hardest negatives \cite{pinsage, eclare}; and ad-hoc techniques like similarity search \cite{slice}, stratified sampling \cite{sampling_cf_nn}, \etc On the other hand, sampling is also generally employed for evaluating recommendation algorithms by estimating expensive to compute metrics like Recall, nDCG, \etc \cite{sampled_metrics}. Finally, sampling is also used to create smaller sub-samples of the \emph{entire} data for reasons like fast experimentation, benchmarking different algorithms, privacy concerns, \etc However, the consequences of different sampling strategies on any of these downstream applications is under-studied, and is the main research interest of this paper. 

\paragraph{Coreset selection.} Closest to our work, a coreset is loosely defined to be a subset of the data-points that maintain a similar ``quality'' as the full dataset for subsequent model training. Submodular approaches try to optimize a function $f : \mathbf{V} \mapsto \mathcal{R}_+$ which measures the utility of a subset $\mathbf{V} \subseteq \mathbf{X}$, and use these estimated functions as a proxy to select the best data subset \cite{coreset_1}. More recent works treat coreset selection as a bi-level optimization problem \cite{coreset_bilevel, coreset_bilevel_2} and directly optimize for the best possible subset for the downstream task. Selection-via-proxy \cite{svp} is another technique which employs an inexpensive base-model as a proxy to tag the importance of each data-point. Note however that all of the discussed coreset selection approaches were designed primarily for classification data, whereas adapting them for interaction data is non-trivial because of: (1) the inherent data heterogeneity; (2) wide range of metrics to evaluate the utility of a recommendation algorithm; and (3) the prevalent missing-data characteristics in user-item interaction data.


\section{Sampling Collaborative Filtering Datasets} Given our motivation of quickly benchmarking recommendation algorithms, we aim to \emph{characterize} the performance of various sampling strategies. We loosely define the performance of a sampling scheme as it's ability in effectively retaining the performance-ranking of different algorithms on the full \vs sub-sampled data. In this section, we start by discussing the different recommendation settings we consider, along with a representative sample of popular recommendation algorithms that we aim to efficiently benchmark. We then examine some popular data sampling strategies, followed by discussing the specifics of \sampler.

\subsection{Problem Settings \& Methods Compared} \label{feedback_types} \label{algorithms}
To give a representative sample of typical recommendation scenarios, we consider three different user feedback settings. In \emph{explicit feedback}, each user $u$ gives a numerical rating $r^u_i$ to each interacted item $i$; the model must predict these ratings for novel user-item interactions. Models from this class are evaluated in terms of the Mean Squared Error (MSE) of the predicted ratings. In \emph{implicit feedback}, only the positive interactions for each user are available
(\eg clicks or purchases), whilst all non-interacted items are considered as negatives. AUC, Recall@$100$, and nDCG@$10$ metrics are used to evaluate model performance for implicit feedback algorithms. Finally, in \emph{sequential feedback}, each user $u$ interacts with an ordered sequence of items. Given these temporal sequences, the goal is to identify the \emph{next-item} that each user $u$ is most likely to interact with.
We use the same metrics as in implicit feedback. Note that following recent warnings against sampled metrics for evaluating recommendation algorithms \cite{sampled_metrics}, we compute both Recall and nDCG by ranking \emph{all} items in the dataset. Further specifics about the data pre-processing, train/test splits, \etc are discussed in-depth in \cref{main_exp}. 

Given the diversity of the scenarios discussed above, there are numerous relevant recommendation algorithms. We use the following seven algorithms, intended to span the state-of-the-art and standard baselines:
\begin{itemize} \setlength\itemsep{0.1em}
    \listheader{PopRec:} A na\"ive baseline that ranks items according to train-set popularity. Note that this method is 
    user independent,
    and has the \emph{same global ranking} of all items.
    
    \listheader{Bias-only:} Another simple baseline that assumes no 
    cross user-item interactions.
    Formally, it learns a global bias $\alpha$, scalar biases $\beta_u, \beta_i$ for each user $u \in \mathcal{U}$ and 
    each item $i \in \mathcal{I}$. Ultimately, the rating/relevance for user $u$ and item $i$ is modeled as $\hat{r}^u_i = \alpha + \beta_u + \beta_i$.

    \listheader{Matrix Factorization (MF) \cite{mf}:} Represents both users and items in a shared latent-space by factorizing the user-item interaction matrix. Formally, the rating/relevance for user $u$ and item $i$ is modeled as $\hat{r}^u_i = \alpha + \beta_u + \beta_i + \gamma_u \cdot \gamma_i$ where $\gamma_u, \gamma_i \in \mathbb{R}^d$ are learned latent representations. 
    
    \listheader{Neural Matrix Factorization (NeuMF) \cite{neural_mf}:} Leverages the representation power of neural networks to capture high level correlations between user and item embeddings. Formally, the rating/relevance for user $u$ and item $i$ is modeled as $\hat{r}^u_i = \alpha + \beta_u + \beta_i + f(\gamma_u ~||~ \gamma_i ~||~ \gamma_u \cdot \gamma_i)$ where $\gamma_u, \gamma_i \in \mathbb{R}^d$, `$||$' is the concatenation operation, and $f : \mathbb{R}^{3d} \mapsto \mathbb{R}$ represents an arbitrarily complex neural network. 
    
    \listheader{Variational Auto-Encoders for Collaborative Filtering (MVAE) \cite{mvae}:} Builds upon the VAE \cite{vae} framework to learn a low-dimensional representation of a user's entire consumption history. More specifically, MVAE encodes each user's bag-of-words consumption history using a VAE and further decodes the latent representation to obtain the completed user preference over all items.
    
    \listheader{Sequential Variational Auto-Encoders for Collaborative Filtering (SVAE) \cite{svae}:} Sequential recommendation algorithm that combines the temporal modeling capabilities of a Gated Recurrent Unit (GRU) \cite{gru} along with the representation power of VAEs. Unlike MVAE, SVAE uses a GRU to encode the user's consumption sequence followed by a multinomial VAE at teach time-step to model the likelihood of the next-item. 
    
    \listheader{Self-attentive Sequential Recommendation (SASRec) \cite{sasrec}:} Another sequential recommendation algorithm that relies on the sequence modeling capabilities of self-attentive neural networks \cite{self_attention}.
    To be precise, given a user $u$ and it's time-ordered consumption history  $\mathcal{S}^u = (\mathcal{S}^u_1, \mathcal{S}^u_2, \ldots, \mathcal{S}^u_{|\mathcal{S}^u|})$, SASRec first applies self-attention on $\mathcal{S}^u$ followed by a series of non-linear feed-forward layers to finally obtain the next-item likelihood.
\end{itemize}
Since Bias-only, MF, and NeuMF can be trained for all three CF-scenarios, we optimize them using the regularized least-squares regression loss for the explicit feedback task, and the pairwise-ranking (BPR \cite{bpr}) loss for implicit/sequential feedback tasks. Note however that the aforementioned algorithms are only intended to be a representative sample of a wide-pool of recommendation algorithms, and in our pursuit to benchmark recommender system models faster, we are primarily concerned with the \emph{ranking} of these algorithms on the full dataset \vs a smaller sub-sample.

\subsection{Sampling Strategies} \label{common_sampling_schemes}
Given a user-item CF dataset \dataset, we aim to create a $p\%$ subset $\mathcal{D}^{s, p}$ according to some sampling strategy $s$. To ensure a fair comparison amongst the different kinds of sampling schemes, we retain exactly $p\%$ of the \emph{total interactions} in $\mathcal{D}^{s, p}$. In this paper, to be comprehensive, we consider eight popular sampling strategies, which can be grouped into the following three categories:

\vspace{-0.2cm}
\paragraph{Interaction sampling.} We first discuss three strategies that sample interactions from \dataset. In \emph{Random Interaction Sampling}, we generate $\mathcal{D}^{s, p}$ by randomly sampling $p\%$ of all the user-item interactions in \dataset. \emph{User-history Stratified Sampling} is another popular sampling technique (see \eg \cite{svae, handbook}) which matches the user-frequency distribution amongst \dataset and $\mathcal{D}^{s, p}$ by randomly sampling $p\%$ of interactions from each user's consumption history in \dataset. In contrast to user-history stratified sampling, \emph{User-history Temporal Sampling} samples $p\%$ of the \emph{most recent} interactions for each user. This strategy is representative of the popular practice of making data subsets from the online traffic of the last $x$ days \cite{pfastre}.

\vspace{-0.2cm}
\paragraph{User sampling.} We also consider two strategies which sample users in \dataset. 
In \emph{Random User Sampling}, we retain users from \dataset at random. More specifically, we iteratively preserve \emph{all} the interactions for a random user until we have retained $p\%$ of the original interactions. In \emph{Head User Sampling}, we iteratively remove the user with the least number of interactions. This method is representative of commonly used pre-processing strategies (see \eg \cite{mvae, neural_mf}) to make data suitable for parameter-heavy algorithms. Such sampling
could be inherently biased
toward users from minority groups and raise concerns from a fairness and diversity perspective \cite{fairness}.

\vspace{-0.2cm}
\paragraph{Graph sampling.} Instead of sampling directly from \dataset, we also consider three strategies that sample from the underlying user-item bipartite interaction graph $\mathcal{G}$. In \emph{Centrality-based Sampling}, we proceed by computing the pagerank centrality scores \cite{pagerank} for each node in $\mathcal{G}$, and retain all the edges (interactions) of the \emph{top scoring nodes} until a total $p\%$ of the original interactions have been preserved. In \emph{Random-walk Sampling} \cite{large_graphs}, we perform multiple random-walks with restart on $\mathcal{G}$ and retain the edges amongst those pairs of nodes that have been visited at least once. We keep expanding our walk until $p\%$ of the initial edges have been retained. We also utilize \emph{Forest-fire Sampling} \cite{forest_fire}, which is a snowball sampling method and proceeds by randomly ``burning'' the outgoing edges of visited nodes. It initially starts with a random node, and then propagates to a random subset of previously unvisited neighbors. The propagation is terminated once we have created a graph-subset with $p\%$ of the initial edges.

\subsection{\emph{\sampler}: Selection-Via-Proxy for CF data} \label{svp_cf} 
Selection-Via-Proxy (SVP) \cite{svp} is a leading coreset mining technique for multi-class classification scenarios. The main idea proposed was simple and effective, and proceeded by training an inexpensive base-model as a proxy to tag the ``importance'' of a data-point. However, applying SVP to CF-data can be highly non-trivial because of the following impediments:
\begin{itemize} \setlength\itemsep{0.1em}
    \listheader{Data heterogeneity:} Unlike classification data $\mathcal{D}_c = \left\{ (x, y) ~|~ x \in \mathbb{R}^d, y\in \mathcal{Y}\right\}$ over some label-space $\mathcal{Y}$, CF-data consists of numerous four-tuples $\{u, i, r^u_i, t^u_i\}$. Such multimodal data adds a lot of different dimensions to sample data from, making it increasingly complex to define meaningful samplers. 
    
    \listheader{Defining the importance of a data-point:} Unlike classification, where we can measure the performance of a classifier by it's empirical risk on held-out data, for recommendation there are a variety of different scenarios (\cref{feedback_types}) along with a wide list of relevant metrics to evaluate the performance of an algorithm. To this effect, it becomes challenging to adapt importance-tagging techniques like greedy k-centers \cite{k_centers}, forgetting-events \cite{forgetting_events}, \etc for recommendation tasks.
    
    \listheader{Missing data:} 
    CF-data is well-known for it's sparsity, skewed and long-tail user/item distributions, and missing-not-at-random (MNAR) characteristics. This results in additional problems as we are now sampling data from skewed, MNAR data, especially using proxy-models trained on the same skewed data. Such sampling in the worst-case might lead to exacerbating existing biases in the data or even aberrant data samples.
\end{itemize}
To address these fundamental limitations in applying the SVP philosophy to CF-data, we propose \sampler to sample representative subsets from large user-item interaction data. \sampler is also specifically devised for our objective of benchmarking different recommendation algorithms, as it relies on the crucial assumption that the ``easiest'' part of a dataset will generally be easy \emph{for all} algorithms. Under this assumption, even after removing such \emph{easy} data we are still likely to retain the overall algorithms' ranking.

Because of the inherent data heterogeneity in user-item interaction data, we can sub-sample in a variety of different ways. We design \sampler to be versatile in this aspect as it can be applied to sample users, items, interactions, or combinations of them, by marginally adjusting the definition of importance of each data-point. In this paper, we limit the discussion to only sampling users and interactions (separately), but extending \sampler for sampling across other data modalities should be relatively straightforward.

Irrespective of whether to sample users or interactions, \sampler proceeds by training an inexpensive proxy model $\mathcal{P}$ on the full, original data \dataset and modifies the forgetting-events approach \cite{forgetting_events} to retain the points with the \emph{highest} importance. More specifically, for explicit feedback, we define the importance of each data-point as $\mathcal{P}$'s average MSE (over epochs) of the specific interaction if we're sampling interactions \emph{or} $\mathcal{P}$'s average MSE of the user (over epochs) if we're sampling users. Whereas, for implicit and sequential feedback, we use $\mathcal{P}$'s average inverse-AUC while computing the importance of each data-point. To be comprehensive, we experiment with both Bias-only and MF as two different proxy-models for \sampler. 

Ultimately, to handle the missing-data and long-tail problems, we also propose \samplerprop which employs user and item propensities to correct the distribution mismatch while estimating the importance of each datapoint. More specifically, let $p_{u, i} = P(r^u_i = 1 ~|~ \overstar{r}^u_i = 1)$ denote the probability of user $u$ and item $i$'s interaction actually being observed (propensity), $E$ be the total number of epochs that $\mathcal{P}$ was trained for, $\mathcal{P}_e$ denote the proxy model after the $e^{\mathit{th}}$ epoch, $\mathcal{I}_u^+ \coloneqq \{ j ~|~ r^u_j > 0 \}$ denote the set of positive interactions for $u$, and $\mathcal{I}_u^- \coloneqq \{ j ~|~ r^u_j = 0 \}$ denote the set of negative interactions for $u$; then, the importance function for \samplerprop, $\mathcal{I}_p$ is defined as follows:
\begin{equation*}
\begin{gathered}
    \mathcal{I}_p(u ~|~ \mathcal{P}) \coloneqq \frac{1}{|\mathcal{I}_u^+|} \cdot \sum_{i \in \mathcal{I}_u^+} \mathcal{I}_p(u, i ~|~ \mathcal{P}) \\
    \mathcal{I}_p(u, i ~|~ \mathcal{P}) \coloneqq \frac{\Delta(u, i ~|~ \mathcal{P})}{p_{u, i}} \\
    \Delta(u, i ~|~ \mathcal{P}) \coloneqq 
    \begin{dcases} 
      ~~ \sum_{e=1}^{E} \left(\mathcal{P}_e(u, i) - r^u_i\right)^2 \\ 
      ~~ \text{(for explicit feedback)} \\ \vspace{1cm}
      ~~ \sum_{e=1}^{E} \sum_{j \sim \mathcal{I}_u^-} \frac{1}{\mathds{1}\left(\mathcal{P}_e(u, i) > \mathcal{P}_e(u, j)\right)} \\
      ~~ \text{(for implicit/sequential feedback)}
  \end{dcases}
\end{gathered}
\end{equation*}

\begin{customTheorem}{3.1}
Given an ideal propensity-model $p_{u, i}; \ \mathcal{I}_p(u, i ~|~ \mathcal{P})$ is an unbiased estimator of $\Delta(u, i ~|~ \mathcal{P})$.
\end{customTheorem}

\begin{proof}
Please check \cref{appendix_derivations}.
\end{proof} 

\paragraph{Propensity model.} A wide variety of ways exist to model the propensity-score of a user-item interaction \cite{propensity_3, pfastre, sachdeva_kdd20}. The most common ways comprise using machine learning models like na\"ive bayes and logistic regression, or by fitting handcrafted, parametrized functions. For our problem statement, we make a simplifying assumption that the data noise is one-sided \ie $P(r^u_i = 1 ~|~ \overstar{r}^u_i = 0)$ or the probability of a user interacting with a \emph{wrong} item is \emph{zero}, and model the probability of an interaction going missing to decompose over the user and item as follows:
\begin{align*}
    p_{u, i} &= P(r^u_i = 1 ~|~ \overstar{r}^u_i = 1) \\
    &= P(r^u = 1 ~|~ \overstar{r}^u = 1) \cdot P(r_i = 1 ~|~ \overstar{r}_i = 1) ~=~ p_u \cdot p_i
\end{align*}
Following \cite{pfastre}, we assume the user and item propensities to lie on the following sigmoid curves:
\begin{align*}
\begin{split}
    p_u \coloneqq \frac{1}{1 + C_u \cdot e^{-A \cdot log(N_u + B)}}
    \\
    p_i \coloneqq \frac{1}{1 + C_i \cdot e^{-A \cdot log(N_i + B)}}
\end{split}
\end{align*}
Where, $N_u$ and $N_i$ represent the total number of interactions of user $u$ and item $i$ respectively, $A$ and $B$ are two fixed scalars, $C_u = (log(|\mathcal{U}|) - 1) \cdot (B+1)^A$ and $C_i = (log(|\mathcal{I}|) - 1) \cdot (B+1)^A$. 

\subsection{Measuring the performance of a sampling strategy} \label{sampling_perf}
To quantify the performance of a sampling strategy $s$ on a dataset $\mathcal{D}$, we start by creating various $p \in \{ 80, 60, 40, 20, 10, 1 \}\%$ subsets of $\mathcal{D}$ according to $s$ and call them $\mathcal{D}^{s, p}$. Next, we train and evaluate all the relevant recommendation algorithms on both $\mathcal{D}$ and $\mathcal{D}^{s, p}$. Let the \emph{ranking} of all algorithms according to CF-scenario $f$ and metric $m$ trained on $\mathcal{D}$ and $\mathcal{D}^{s, p}$ be $\mathcal{R}_{f, m}$ and $\mathcal{R}^{s, p}_{f, m}$ respectively, then the performance measure $\Psi(\mathcal{D}, s)$ is defined as the average correlation between $\mathcal{R}_{f, m}$ and $\mathcal{R}^{s, p}_{f, m}$ measured through Kendall's Tau over all possible CF-scenarios, metrics, and sampling percents:
\begin{equation*}
\vspace{-0.05cm}
    \Psi\left(\mathcal{D}, s\right) = \lambda \cdot \sum_{f} \sum_{m} \sum_{p} \tau\left(\mathcal{R}_{f, m}, \mathcal{R}^{s, p}_{f, m}\right)
\vspace{-0.15cm}
\end{equation*}
Where $\lambda$ is an appropriate normalizing constant for computing the average.
$\Psi$ has the same range as Kendall's Tau \ie $[-1, 1]$ and a higher $\Psi-$value indicates strong agreement between the algorithm ranking on the full and sub-sampled datasets, whereas a large negative $\Psi-$value implies that the algorithm order was effectively reversed.

\section{Experiments} \label{main_exp}

\paragraph{Datasets.} We use six public CF datasets with varying sizes, sparsity patterns, \etc We use three different subsets (Magazine, Luxury, and Video-games) of the Amazon review datasets \cite{amz_data}, along with the Movielens-100k \cite{movielens}, BeerAdvocate \cite{beer_dataset}, and GoodReads Comics \cite{mengting_goodreads} datasets. 
We simulate all three feedback scenarios (\cref{feedback_types}) for each dataset via different pre-processing 
strategies. For explicit and implicit feedback, we follow a randomized 80/10/10 train-test-validation split for each user's consumption history, and make use of the leave-one-last \cite{train_test_splitting} strategy for the sequential feedback task. 
In pursuit of following the least restrictive data pre-processing \cite{making_progress}, we only weed out the users that have lesser than 3 interactions in total.
Please check \cref{appendix_exp_details} for further information about data statistics and training details.

\paragraph{How do different sampling strategies compare to each other?} $\Psi$-values for all sampling schemes on all datasets can be found in \cref{psi_results}. Even though there are only six datasets under consideration, there are a few prominent patterns. First, the average $\Psi$ for most sampling schemes is around $0.4$, which implies a statistically significant correlation between the ranking of algorithms on the full \vs sub-sampled datasets. Next, \sampler generally outperforms all commonly used sampling strategies by some margin in retaining the ranking of different recommendation algorithms. Finally, strategies that discard the tail-part of a dataset (head-user, centrality-based) are the worst performing overall, which supports the recent warnings against dense sampling of data \cite{sigir20}.

\newcommand{\STAB}[1]{\begin{tabular}{@{}c@{}}#1\end{tabular}}
\begin{table*}
    \begin{scriptsize} 
    \begin{center}
        \begin{tabular}{c c | c c c c c c | c}
            \toprule
            \multicolumn{2}{c|}{\multirow{4}{*}{\textbf{Sampling strategy}}} & \multicolumn{6}{c|}{\emph{Datasets}} & \\
            & & \multicolumn{6}{c|}{} & \\
            & & \begin{tabular}{@{}c@{}}\textbf{Amazon}\\\textbf{Magazine}\end{tabular} & \textbf{ML-100k} & \begin{tabular}{@{}c@{}}\textbf{Amazon}\\\textbf{Luxury}\end{tabular} & \begin{tabular}{@{}c@{}}\textbf{Amazon}\\\textbf{Video-games}\end{tabular} & \begin{tabular}{@{}c@{}}\textbf{Beer}\\\textbf{Advocate}\end{tabular} & \begin{tabular}{@{}c@{}}\textbf{Goodreads}\\\textbf{Comics}\end{tabular} & \textbf{\emph{Average}} \\
            \midrule
            
            \multirow{8}{*}{\STAB{\rotatebox[origin=c]{90}{\begin{tabular}{@{}c@{}}Interaction sampling\\\end{tabular}}}} & Random & 0.428     &  0.551     &  0.409     &  0.047     &  0.455     &  0.552     &  0.407 \\[0.6mm]
            & Stratified & 0.27      &  0.499     &  0.291     &  -0.01     &  0.468     &  0.538     &  0.343 \\[0.6mm]
            & Temporal & 0.289     &  0.569     &  0.416     &  -0.02     &  \underline{0.539}     &  0.634     &  0.405 \\[0.6mm]
            & \sampler \emph{w/} MF & 0.418     &  0.674     &  0.398     &  0.326     &  0.425     &  \underline{0.662}     &  \underline{0.484} \\[0.6mm]
            & \sampler \emph{w/} Bias-only & 0.38      &  0.684     &  \underline{0.431}     &  \underline{0.348}     &  0.365     &  0.6       &  0.468 \\[0.6mm]
            & \samplerprop \emph{w/} MF & 0.381     &  0.617     &  0.313     &  0.305     &  0.356     &  0.608     &  0.43 \\[0.6mm]
            & \samplerprop \emph{w/} Bias-only & 0.408     &  0.617     &  0.351     &  0.316     &  0.437     &  0.617     &  0.458 \\[0.6mm]
            \midrule
            \multirow{7}{*}{\STAB{\rotatebox[origin=c]{90}{\begin{tabular}{@{}c@{}}User sampling\\\end{tabular}}}} & Random & 0.436     &  0.622     &  0.429     &  0.17      &  0.344     &  0.582     &  0.431 \\[0.6mm]
            & Head & 0.369     &  0.403     &  0.315     &  0.11      &  -0.04     &  -0.02     &  0.19 \\[0.6mm]
            & \sampler \emph{w/} MF & 0.468     &  0.578     &  0.308     &  0.13      &  0.136     &  0.444     &  0.344 \\[0.6mm]
            & \sampler \emph{w/} Bias-only & 0.49      &  0.608     &  0.276     &  0.124     &  0.196     &  0.362     &  0.343 \\[0.6mm]
            & \samplerprop \emph{w/} MF & 0.438 &  0.683 &  0.307 &  0.098 &  0.458 &  0.592 &  0.429 \\[0.6mm]
            & \samplerprop \emph{w/} Bias-only & 0.434     &  \underline{0.751}     &  0.233     &  0.107     &  0.506     &  0.637     &  0.445 \\[0.6mm]
            \midrule
            \multirow{4}{*}{\STAB{\rotatebox[origin=c]{90}{\begin{tabular}{@{}c@{}}Graph\\\end{tabular}}}} & Centrality & 0.307     &  0.464     &  0.407     &  0.063     &  0.011     &  0.343     &  0.266 \\[0.6mm]
            & Random-walk & \underline{0.596}  &  0.5       &  0.395     &  0.306     &  0.137     &  0.442     &  0.396 \\[0.6mm]
            & Forest-fire & 0.564  &  0.493   &  0.415   &  0.265   &  0.099  &  0.454  &  0.382 \\[0.6mm]
            \bottomrule
        \end{tabular}
    \end{center}
    \end{scriptsize}
    \caption{$\Psi$-values for all datasets and sampling strategies. Higher $\Psi$ is better. The best $\Psi$ for every dataset is \underline{underlined}. The $\Psi$-values for each sampling scheme \emph{averaged over all datasets} is appended to the right.}
    \label{psi_results}
\end{table*}

\begin{figure}[ht!] 
    \centering
    \includegraphics[width=0.7\linewidth]{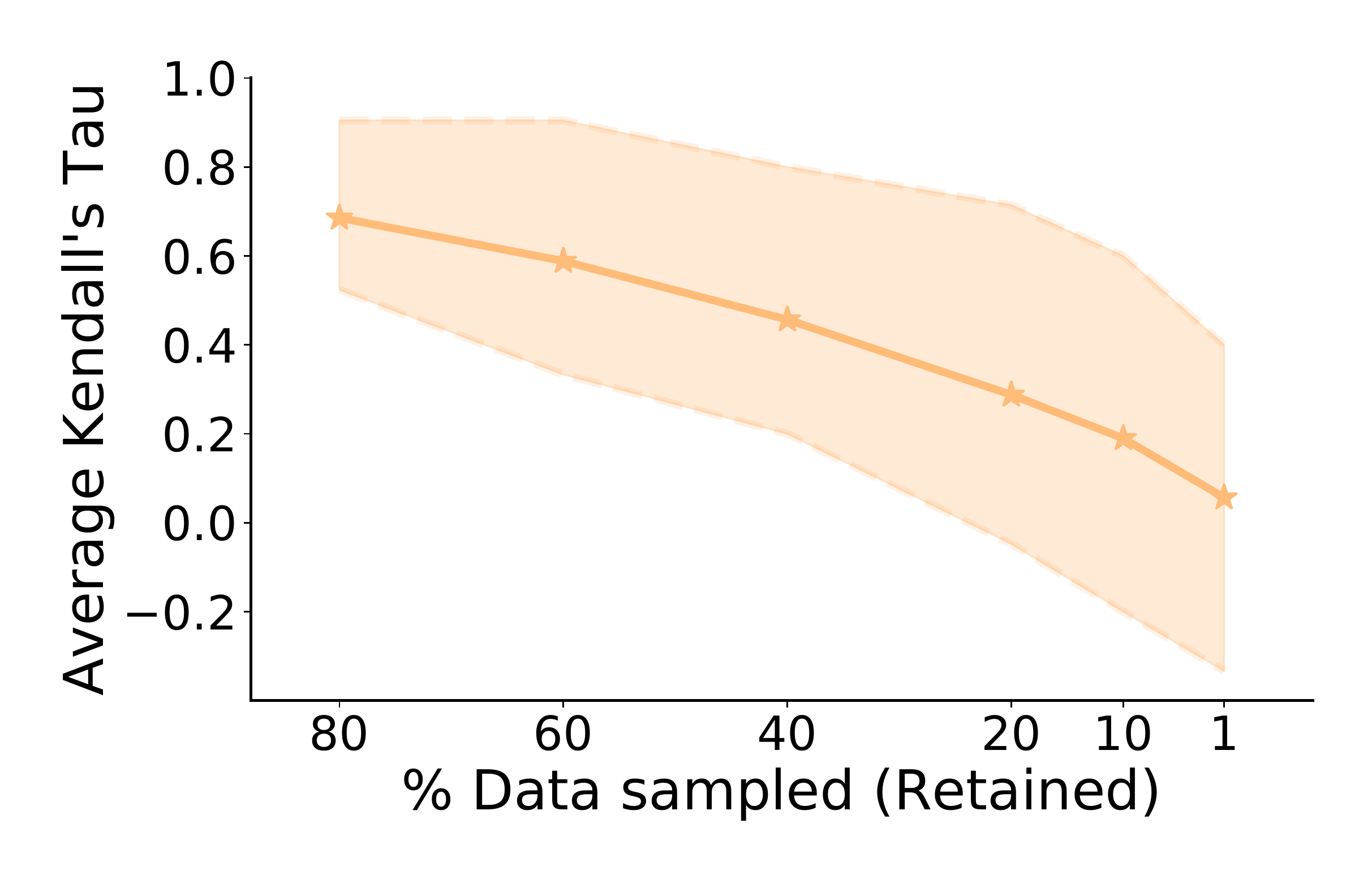}
    \caption{Comparison of the average Kendall's Tau with \% data sampled. A higher Tau indicates better retaining power.}
    \label{percent_sampling_vs_tau}
\end{figure}

\paragraph{How much data to sample?} Since $\Psi$ is averaged over all $p \in \{ 80, 60, 40, 20, 10, 1 \}$\% data samples, to better understand a reasonable amount of data to sample, we stratify $\Psi$ according to each value of $p$ and note the average Kendall's Tau. As we observe from \cref{percent_sampling_vs_tau}, there is a steady increase in the performance measure when more data is retained. Next, despite the results being averaged over \emph{sixteen} different sampling strategies, $50-60\%$ of the data seems enough for gauging the algorithm order.

\begin{figure}[ht!] 
    \centering
    \includegraphics[width=\linewidth]{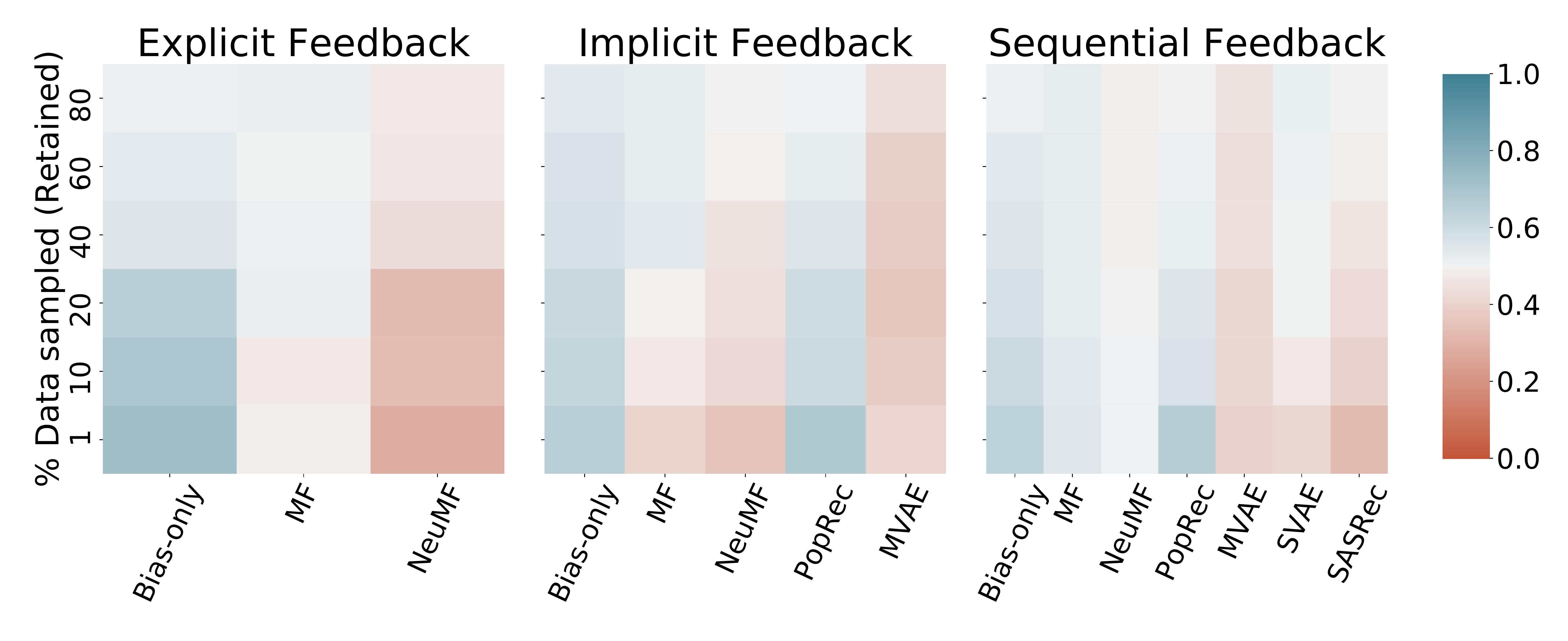}
    \vspace{-0.6cm}
    \caption{Heatmap of $P_{\mathit{MLE}}$. A deep blue indicates that the algorithm is most probable to \emph{move up} in the sampled data ranking order, whereas a deep red indicates that the algorithm is most probable to \emph{move down}.}
    \label{percent_sampling_vs_method}
\end{figure}

\paragraph{How does the relative performance of algorithms change as a function of sampling rate?} In an attempt to better understand the impact of sampling on different recommendation algorithms used in this study (\cref{algorithms}), we visualize the probability of an algorithm moving in the overall method ranking with data sampling. We estimate the aforementioned probability using Maximum-Likelihood-Estimation (MLE) on the experiments run in computing $\Psi(\mathcal{D}, s)$. Formally, given a recommendation algorithm $r$, CF-scenario $f$, and data sampling percent $p$:
\begin{equation*}
    P_{\mathit{MLE}}(r ~|~ f, p) = \lambda \cdot \sum_{\mathcal{D}} \sum_{s} \sum_{m} 0.5 + \frac{\mathcal{R}_{f, m}(r) - \mathcal{R}_{f, m}^{s, p}(r)}{2 \cdot (n-1)}
\end{equation*}
where $\lambda$ is an appropriate normalizing constant, and $n$ represents the total number of recommendation algorithms. A heatmap visualizing $P_{\mathit{MLE}}$ for all recommendation algorithms and CF-scenarios is shown in \cref{percent_sampling_vs_method}. We observe that simpler methods like Bias-only and PopRec are most probable to move upwards in the ranking order with extreme sampling whereas parameter-heavy algorithms like SASRec, SVAE, MVAE, \etc tend to move downwards.

\begin{figure}[ht!] 
    \centering
    \includegraphics[width=\linewidth]{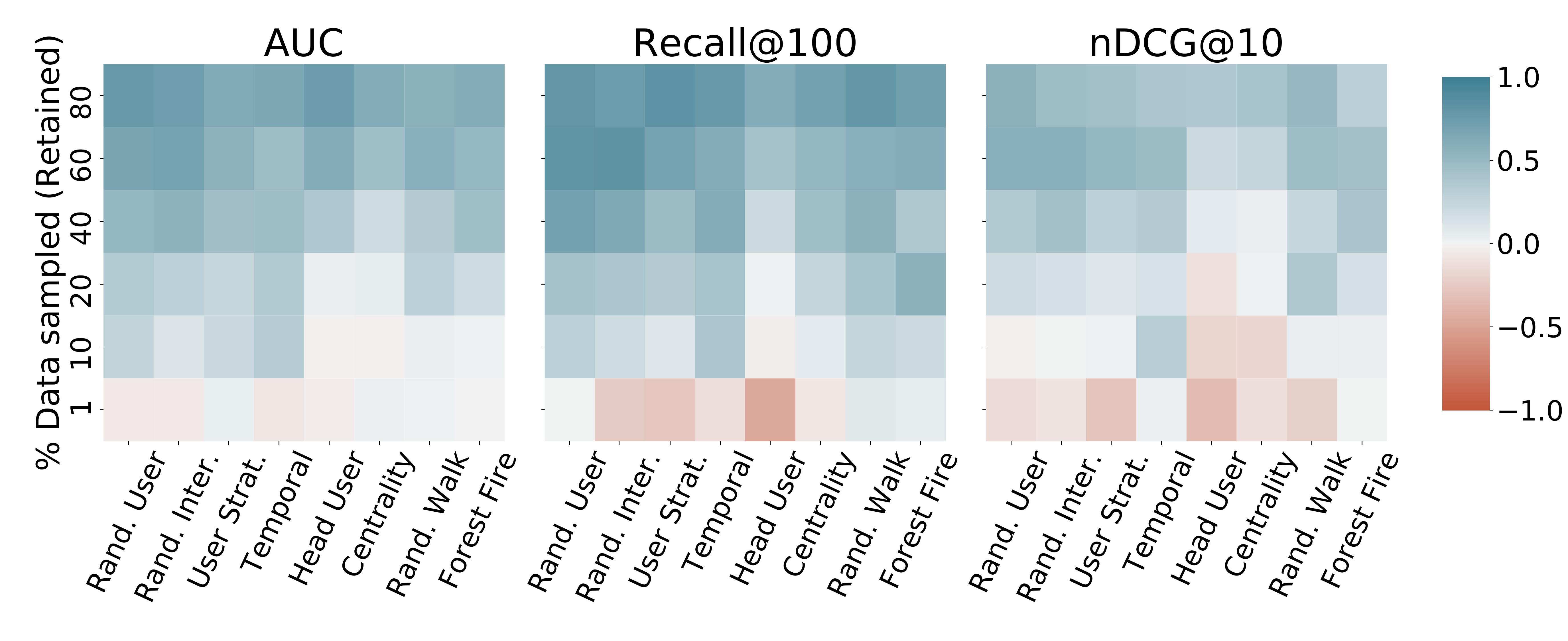}
    \vspace{-0.6cm}
    \caption{Heatmap of the average Kendall's Tau for different sampling strategies stratified over metrics and \% data sampled.}
    \label{metric_correlation}
\end{figure}

\paragraph{Are different metrics affected equally by sampling?} To better understand how the different implicit and sequential feedback metrics (\cref{feedback_types}) are affected by sampling, we visualize the average Kendall's Tau for all sampling strategies (except \sampler for brevity) and all \% data sampling choices separately over 
each metric
in \cref{metric_correlation}. We observe a steady decrease in the model quality across accuracy metrics and sampling schemes. This is in agreement with the analysis from \cref{percent_sampling_vs_tau}. Next, most sampling schemes follow a \emph{similar} downwards trend in performance for all three metrics with AUC being slightly less and nDCG being slightly more affected by sampling.

\section{Conclusion} In this work, we characterized the performance of various sampling strategies for the task of accurately retaining the \emph{relative} performance of different recommendation algorithms. We also proposed a novel sampling scheme, \sampler, which is better than commonly used strategies and can confidently gauge the best performing algorithm with only half of the initial dataset. An interesting research direction in addition to more representative sampling of CF-data is analyzing the fairness and privacy aspects of training algorithms on sub-sampled data, that we delay for future work.

\bibliography{main}
\bibliographystyle{icml2021}


\appendix
\onecolumn

\section{Derivations} \label{appendix_derivations} \begin{customTheorem}{3.1}
Given an ideal propensity-model $p_{u, i}; \ \mathcal{I}_p(u, i ~|~ \mathcal{P})$ is an unbiased estimator of $\Delta(u, i ~|~ \mathcal{P})$.
\end{customTheorem}

\begin{proof}
\begin{align*}
    \EE_{u \sim \mathcal{U}} \EE_{i \sim \mathcal{I}} \left[ \mathcal{I}_p(u, i ~|~ \mathcal{P}) \right] &= \frac{1}{|\mathcal{U}| \cdot |\mathcal{I}|} \sum_{u \sim \mathcal{U}} \sum_{i \sim \mathcal{I}} \mathcal{I}_p(u, i ~|~ \mathcal{P}) \cdot P(r_i^u = 1) \\
    &= \frac{1}{|\mathcal{U}| \cdot |\mathcal{I}|} \sum_{u \sim \mathcal{U}} \sum_{i \sim \mathcal{I}} \mathcal{I}_p(u, i ~|~ \mathcal{P}) \cdot \left( P(r_i^u = 1, \overstar{r}_i^u = 0) ~+~ P(r_i^u = 1, \overstar{r}_i^u = 1) \right) \\
    &= \begin{aligned}[t]
        \frac{1}{|\mathcal{U}| \cdot |\mathcal{I}|} \sum_{u \sim \mathcal{U}} \sum_{i \sim \mathcal{I}} \frac{\Delta(u, i ~|~ \mathcal{P})}{p_{u, i}} \cdot ( P(\overstar{r}_i^u = 0) \cdot \cancelto{0}{P(r_i^u = 1 ~|~ \overstar{r}_i^u = 0)} \hspace{1cm} \text{(One-sided label noise)} \\
        +~ P(\overstar{r}_i^u = 1) \cdot P(r_i^u = 1 ~|~ \overstar{r}_i^u = 1) ) \hspace{4cm} \\
    \end{aligned} \\
    &= \frac{1}{|\mathcal{U}| \cdot |\mathcal{I}|} \sum_{u \sim \mathcal{U}} \sum_{i \sim \mathcal{I}} \Delta(u, i ~|~ \mathcal{P}) \cdot P(\overstar{r}_i^u = 1) \\ 
    &= \EE_{u \sim \mathcal{U}} \EE_{i \sim \mathcal{I}} \left[ \Delta(u, i ~|~ \mathcal{P}) \right] \qedhere
\end{align*}
\end{proof}
\section{Experiment Details} \label{appendix_exp_details} In this section, we provide further details about the experiments in Section 4 of the paper. First, all the collaborative filtering data scenarios that we consider in the paper along with their pertinence with each recommendation algorithm and evaluation metric is listed in \cref{model_scenario_table}. A brief set of data statistics of the six datasets we use is also presented in \cref{data_stats}.

\begin{table*}[!ht]
    \begin{small} 
    \begin{center}
                
                
        \begin{tabular}{c | c c c c c c c | c c c c}
            \toprule
            \multirow{3}{*}{CF-scenario} & \multicolumn{7}{c|}{\emph{Algorithm}} & \multicolumn{4}{c}{\emph{Metric}} \\
            & \multicolumn{7}{c|}{} & \multicolumn{4}{c}{} \\
            & Bias-only & MF & NeuMF & PopRec & MVAE & SVAE & SASRec & MSE & AUC & Recall@k & nDCG@k \\ \midrule
            Explicit & Yes & Yes & Yes & $\times$ & $\times$ & $\times$ & $\times$ & Yes & $\times$ & $\times$ & $\times$ \\[0.6mm]
            Implicit & Yes & Yes & Yes & Yes & Yes & $\times$ & $\times$ & $\times$ & Yes & Yes & Yes \\[0.6mm]
            Sequential & Yes & Yes & Yes & Yes & Yes & Yes & Yes & $\times$ & Yes & Yes & Yes \\[0.6mm]
            \bottomrule
        \end{tabular}
    \end{center}
    \end{small}
    \caption{Demonstrates the pertinence of each CF-scenario towards each recommendation algorithm (left) and each metric (right).}
    \label{model_scenario_table}
\end{table*}

\begin{table}[!ht]
    \begin{footnotesize} 
    \begin{center}
        \begin{tabular}{c | c c c c}
            \toprule
            \textbf{Dataset} & \textbf{\# Interactions} & \textbf{\# Users} & \textbf{\# Items} & \textbf{Avg. \# User Interactions} \\ 
            \midrule
            
            Amazon Magazine      & 12.7k & 3.1k  & 1.3k  & 4.1 \\
            ML-100k              & 100k  & 943   & 1.7k  & 106.04 \\
            Amazon Luxury        & 126k  & 29.7k & 8.4k  & 4.26 \\
            Amazon Video-games   & 973k  & 181k  & 55.3k & 5.37 \\
            BeerAdvocate         & 1.51M & 18.6k & 64.3k & 81.45 \\
            Goodreads     & 4.37M & 133k  & 89k   & 32.72 \\
            
            \bottomrule
        \end{tabular}
    \end{center}
    \end{footnotesize}
    \caption{Data statistics of the \emph{six} datasets used in this paper.}
    \label{data_stats}
\end{table}

\paragraph{Training details.} We implement all algorithms in PyTorch \footnote{All code and data splits will be released at publication time.} and train on a single GTX 1080-Ti GPU. For a fair comparison across recommendation algorithms, we perform a generous hyper-parameter search. For the three smallest datasets used in this paper (\cref{data_stats}), we search the latent-size in $\{ 4, 8, 16, 32, 50 \}$, dropout in $\{ 0.0, 0.3, 0.5 \}$, and the learning rate in $\{ 0.001, 0.006, 0.02 \}$. Whereas for the three largest datasets, we fix the learning rate to be $0.006$ and only search over the latent-size and dropout values. Note that despite the limited number of datasets and recommendation algorithms used in this study, given that we need to train all algorithms with hyper-parameter tuning for all CF-scenarios, $\%$ data sampled according to all different sampling strategies, there are a total of $6 \times 3 \times 7 \times 16 \times 7 \times 30 \approx 400k$ unique models trained.

\paragraph{Data sampling.} To compute the $\Psi$-values of different sampling schemes (Section 3.4), we make six different splits for each sampling scheme with $\{ 80, 60, 40, 20, 10, 1 \} \%$ interactions sampled from the original dataset. To keep comparisons as fair as possible, for all sampling schemes, we only sample on the train-set and never touch the validation and test-sets. This simulates the practical scenario of sampling data only while benchmarking algorithms offline, whereas the live (test-set) traffic still remains the same.
\section{Environmental Impact} \label{appendix_env} \begin{wraptable}{r}{0.4\textwidth}
    \vspace{-4.5mm} 
    \begin{footnotesize} 
    \begin{center}
        \begin{tabular}{c c}
            \toprule
            \textbf{Consumption} & \textbf{CO$_2$e (lbs.)} \\ \midrule
            
            1 person, NY$\leftrightarrow$SF flight      & 2k \\
            Human life, 1 year avg.                     & 11k \\ 
            \midrule
            Weekly RecSys development cycle             & 20k \\
            '' \ \ \ \ \emph{w/} \sampler               & 10.5k \\
            
            \bottomrule
        \end{tabular}
    \end{center}
    \end{footnotesize}
    \caption{CO$_2$ emissions comparison. Source \cite{co2e}}
    \label{co2e}
\end{wraptable}
To realize the real-world environmental impact of \sampler, we discuss a typical weekly RecSys development cycle in the industry and its carbon footprint. Inspired by the Criteo Ad Terabyte dataset, which is a collection of 24 days of ad-click logs, we assume a common industrial dataset to have around $4$ billion interactions; and for a specific use-case want to benchmark for \eg $25$ different algorithms, each with $40$ different hyper-parameter variations. To estimate the run-time and energy efficiency of the GPU hardware, we scale the $0.4$ minute MLPerf \cite{mlperf} run of training NeuMF \cite{neural_mf} on the Movielens-20M dataset over an Nvidia DGX-2 machine. The total estimated run-time for all experiments would be $25 \times 40 \times \frac{4B}{20M} \times \frac{0.4}{60} \approx 1340$ hours; and following \cite{co2e}, the net CO$_2$ emissions would roughly be $10 \times 1340 \times 1.58 \times 0.954 \approx 20k$ lbs. To better understand the significance of this number, a brief CO$_2$ emissions comparison is presented in \cref{co2e}. Clearly, \sampler along with saving a large amount of experimentation time and cloud compute cost, can also significantly reduce the carbon footprint of this \emph{weekly process} by almost an average human's \emph{yearly} CO$_2$ emissions.




\end{document}